\documentclass[aps,pra,twocolumn,amsmath,amssymb,floatfix,superscriptaddress]{revtex4-2}
\usepackage{amsmath}
\usepackage{amssymb}
\usepackage{amsfonts}
\usepackage{bbm}
\usepackage{listings}
\usepackage{enumerate}
\usepackage{latexsym}
\usepackage{multirow}
\usepackage[T1]{fontenc}
\usepackage{graphicx}
\usepackage{braket}
\usepackage[pdftex,colorlinks=false]{hyperref}
\usepackage{xcolor, colortbl}
\usepackage{verbatim}

\usepackage{times}
\usepackage{array, makecell}
\usepackage[export]{adjustbox}
\usepackage{tabularx}
\usepackage{mathtools}
\usepackage{tikz}
\usepackage{siunitx}
\usepackage{filecontents}
\usepackage[normalem]{ulem}

\usepackage{lineno}

\makeatletter
\def\maketitle{
\@author@finish
\title@column\titleblock@produce
\suppressfloats[t]}
\makeatother

\newcolumntype{M}[1]{>{\centering\arraybackslash}m{#1}} 

\newcolumntype{N}{@{}m{0pt}@{}}

\newcommand{\beq}{\begin{equation}}
\newcommand{\eneq}{\end{equation}}

\def\be{\begin{equation}}
\def\ee{\end{equation}}
\def\ba{\begin{eqnarray}}
\def\ea{\end{eqnarray}}

\def\R{{\rm Re}}
\def\Z{\mathbb{Z}}
\def\C{\mathbb{C}}

\def\dag{\dagger}

\def\beq{\begin{equation}}
\def\eeq{\end{equation}}
\def\barray{\begin{eqnarray}}
\def\earray{\end{eqnarray}}

\font\upright=cmu10 scaled\magstep1
\def\stroke{\vrule height8pt width0.4pt depth-0.1pt}

\def\Zmath{\mathbb{Z}}
\def\Qmath{\vcenter{\hbox{\upright\rlap{\rlap{Q}\kern
                   3.8pt\stroke}\phantom{Q}}}}
\def\Nmath{\vcenter{\hbox{\upright\rlap{I}\kern 1.7pt N}}}
\def\Cmath{\vcenter{\hbox{\upright\rlap{\rlap{C}\kern
                   3.8pt\stroke}\phantom{C}}}}
\def\Rmath{\vcenter{\hbox{\upright\rlap{I}\kern 1.7pt R}}}
\def\Z{\ifmmode\Zmath\else$\Zmath$\fi}
\def\Q{\ifmmode\Qmath\else$\Qmath$\fi}
\def\N{\ifmmode\Nmath\else$\Nmath$\fi}
\def\C{\ifmmode\Cmath\else$\Cmath$\fi}
\def\R{\ifmmode\Rmath\else$\Rmath$\fi}

\input{epsf}

\definecolor{MMCOLOR}{RGB}{224, 138, 12}
\newcommand{\MM}[1]{{\textcolor{MMCOLOR}{#1}}}

\newcounter{defcounter}
\newcommand{\affA}{Van der Waals-Zeeman Institute, Institute of Physics,
University of Amsterdam, 1098 XH Amsterdam, Netherlands}
\newcommand{\affB}{QuSoft, Science Park 123, 1098 XG Amsterdam, the Netherlands}
\newcommand{\affC}{Institute for Theoretical Physics, Institute of Physics, University of Amsterdam, Science Park 904, 1098 XH Amsterdam, the Netherlands}

\begin{document}

%
%
%

\title{Trapped Ion Quantum Computing using Optical Tweezers and the Magnus Effect}

\author{M. Mazzanti}\affiliation{\affA}
\author{R. Gerritsma}\affiliation{\affA}\affiliation{\affB}
\author{R. J. C. Spreeuw}\affiliation{\affA}\affiliation{\affB}
\author{A. Safavi-Naini}\affiliation{\affB}\affiliation{\affC}

\begin{abstract}
We consider the implementation of quantum logic gates in trapped ions using tightly focused optical tweezers. Strong polarization gradients near the tweezer focus lead to qubit-state dependent forces on the ion. We show that these may be used to implement quantum logic gates on pairs of ion qubits in a crystal. The qubit-state dependent forces generated by this effect live on the plane perpendicular to the direction of propagation of the laser beams opening new ways of coupling to motional modes of an ion crystal. The proposed gate does not require ground state cooling of the ions and does not rely on the Lamb-Dicke approximation, although the waist of the tightly focused beam needs to be comparable with its wavelength in order to achieve the needed field curvature. Furthermore, the gate can be performed on both ground state and magnetic field insensitive clock state qubits without the need for counter-propagating laser fields. This simplifies the setup and eliminates errors due to phase instabilities between the gate laser beams. Finally, we show that imperfections in the gate execution, in particular pointing errors $<30$~nm in the tweezers reduce the gate fidelity from $
\mathcal F\gtrsim 0.99998$ to $\gtrsim 0.999$.
\end{abstract}

\date{\today}

\maketitle

Trapped ions are one of the most mature platforms for the implementation of quantum computing and quantum logic gates have been implemented with very high fidelity in these systems~\cite{Ballance:2016,Gaebler:2016}. Usually, the quantum logic gates in trapped ions rely on state-dependent forces applied to the ions by laser fields or magnetic fields. The exchange of motional quanta between the ions then leads to effective qubit-qubit interactions. Several recent works have explored how the use of state-of-the-art optical tweezer technology can benefit the trapped ion quantum computer. Optical tweezers can be used to confine atoms very strongly by inducing a dipole in them and find application in neutral atomic quantum simulators, in which tweezers are used to levitate individual atoms~\cite{barredo_atom-by-atom_2016, endres_atom-by-atom_2016, norcia_microscopic_2018, levine_parallel_2019, Browaeys_NP_review}. In trapped ions, tweezers may be used to tune the soundwave spectrum in the ion crystal and thereby to program the interactions between the qubits~\cite{Olsacher:2020,Teoh:2021,Espinoza:2021}.  Furthermore, in a recent work~\cite{Mazzanti:2021} we have proposed combining state-dependent optical tweezers  with oscillating electric fields to build a universal trapped ion quantum computer with extremely long-ranged interactions between the qubits.

In this work, we consider another scenario, in which we make use of the strong polarization gradients that occur in optical tweezers. We note that strong gradients in optical potentials have been previously investigated to implement two-qubit gates without the need for ground-state cooling ~\cite{Cirac:2000, Steanepush, Colarcopush}. However, our approach utilizes the state-dependent displacement of the tweezer potential due to polarization gradients \cite{wang_high-fidelity_2020,spreeuw_off-axis_2020,spreeuw_spiraling_2022}. We propose to use this optical analogue of the Magnus effect to implement quantum logic gates in trapped ions.

\begin{figure}[h!]
    \includegraphics[width=0.48\textwidth]{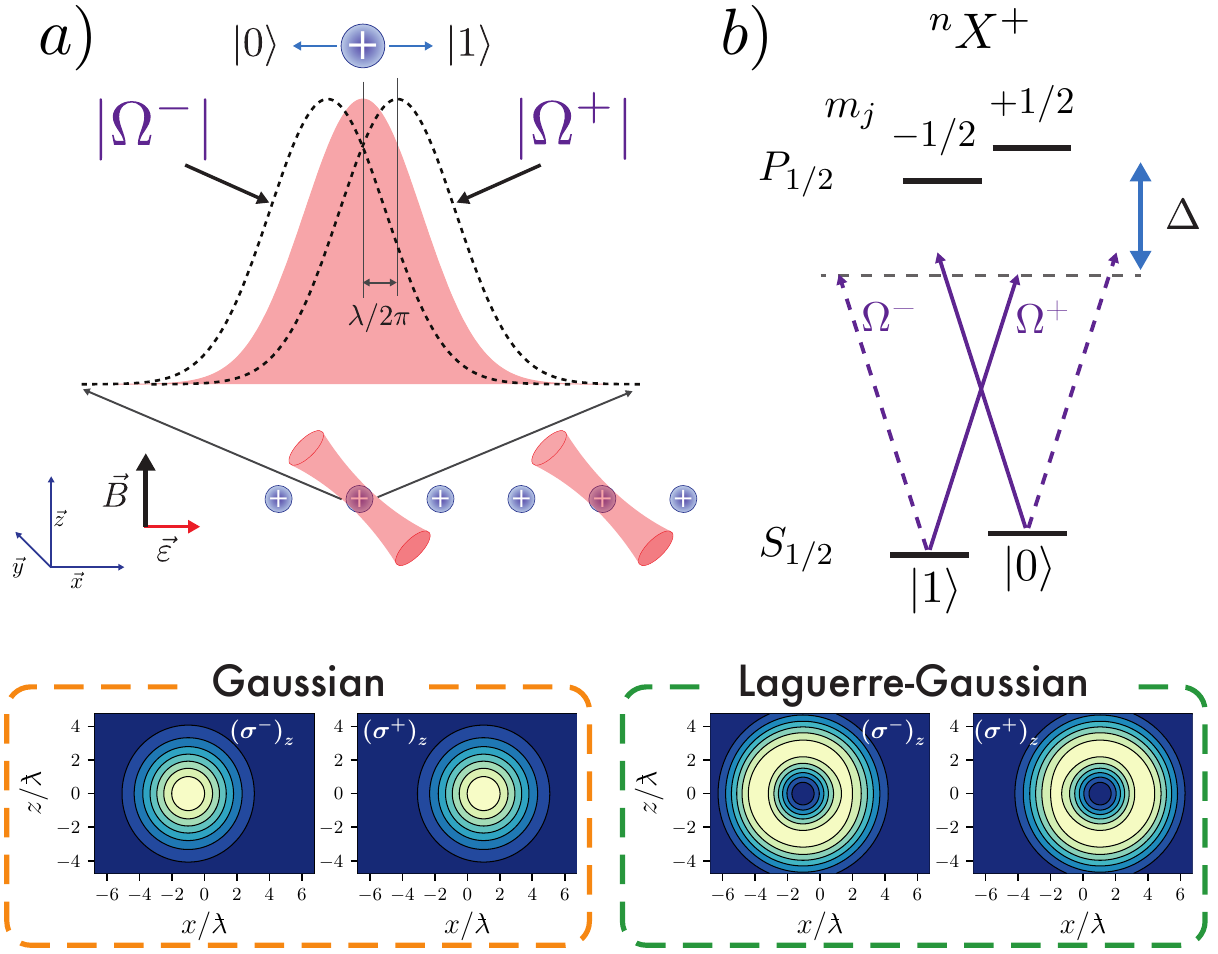}
    \caption{Schematic representation of the two-qubit gate. a) We apply tweezers propagating along the $-y$ direction on the two ions forming the gate. The tweezer intensity can be decomposed into three polarization components. b) Simplified level scheme of an alkaline-earth like ion without nuclear spin showing the encoding of the qubit in its Zeeman ground states. The two polarization components of the tweezer couple to different states in the $P_{1/2}$ manifold with detuning $\Delta$. This causes the minima of the tweezer potentials to be shifted by an amount $\pm\lambdabar$ depending on the qubit state. Bottom : main polarization components for a Gaussian and Laguerre-Gaussian ($l=1$, $n=0$) tightly focused tweezer. }
    \label{fig:174Yb}
\end{figure}

 \emph{Setup --} We consider linearly $x$-polarized, Gaussian tweezers, pointing in the $-y$ direction and tightly focused at two qubits between which we wish to implement a quantum logic gate. The quantum computing platform here considered is a linear crystal of $N$ alkali-like trapped ions of mass $m$. In the focal plane the ions experience a strong polarization gradient along the $x$ direction, such that the polarization is linear ($x$) in the center and circular $(\sigma^\pm)_z$ in the wings of the Gaussian. A direct calculation \cite{Suppl:Material} decomposing the field in the focal plane into its circular components $(\sigma^\pm)_z$ (and $\pi_z$) shows that, to a good approximation, the circular components are near-Gaussian distributions, displaced in opposite directions along the $x$ axis. We depict this setup in Fig.~\ref{fig:174Yb}. Note that the circular components rotate in the $xy$ plane, i.e.\ a plane containing the $\mathbf{k}$ vector of the light. As shown in Fig.\ \ref{fig:174Yb}, the $(\sigma^\pm)_z$ component is displaced by an amount  $\pm\lambdabar\equiv\pm\lambda/2\pi$, with $\lambda$ the tweezer wavelength. As the total field is the superposition of two displaced Gaussians, its intensity is slightly elongated along $x$. Hollow tweezers (Gaussian-Laguerre) can be used instead of Gaussian ones. This will provide the needed field curvature while keeping near-zero intensity at the center of the beam, drastically reducing the probability of off-resonant scattering that might limit the gate fidelity.

For simplicity, we first consider ions without nuclear spin, such as $^{40}$Ca$^+$, $^{88}$Sr$^+$, $^{138}$Ba$^+$ and $^{174}$Yb$^+$. The qubits are encoded in the electronic ground states $^2S_{1/2}$ and $|0\rangle = |j=1/2,m_j=1/2\rangle$ and $|1\rangle = |j=1/2,m_j=-1/2\rangle$ with $j$ the total electronic angular momentum and $m_j$ its projection on the quantization axis. The magnetic field lies along the $z$-direction and the tweezers are polarized along the $x$-direction, such that the ions experience linearly polarized laser light. The direction along the $x$-axis is the long direction of the ion trap, with trap frequency $\omega_x$. The motion of the ions along the $x$-direction can be described by collective modes of harmonic motion with frequencies $\omega_m$ and mode vectors $b_{i,m}$, with $m$ labeling the mode and $i$ the ion~\cite{James:1998}. 

We choose the detuning between the tweezers and the D1 transition to be large enough to avoid photon scattering, but much smaller than the spin-orbit coupling splitting of the $^2P$ state. In this way, we can  neglect coupling to the $P_{3/2}$ state. In what follows we will show that this requirement can be satisfied experimentally. Close to the center of the tweezer, strong polarization gradients appear and as a result, the two qubit states experience slightly different tweezer potentials. In particular, as we show in Fig.~\ref{fig:174Yb}(a), the optical Magnus effect causes each qubit state to experience a tweezer potential that is offset from the apparent center of the tweezer by $\sim\lambdabar$ \cite{spreeuw_off-axis_2020}. Hence, we may approximate the tweezer potential as :
\begin{align}
        \hat U(x)&=-U_0\exp\left(-2(\hat x+\hat{\sigma}_z\lambdabar)^2/w_0^2\right)\\
        &\approx -\tilde{U}_0+\frac{1}{2}m\omega_\text{tw}^2\hat x^2+g x \hat{\sigma_z}
\end{align}
with $\omega_\text{tw}=\sqrt{4\tilde{U}_0(w_0^2-4\lambdabar^2)/(m w_0^4)}$, $g=4\tilde{U}_0\lambdabar/w_0^2$, and $\tilde{U}_0=U_0\exp(-2\lambdabar^2/w_0^2)\approx U_0$. Here $U_0$ is the tweezer potential in the center and the beam waist is $w_0$. 
Our approximation replaces the tweezer potential with a harmonic potential and is valid for $w_0 \gg l_m$, with $l_m = \sqrt{\hbar/2m\omega_m}$. The last term in $U(x)$ is the result of the spin-dependent force $g$ coupling the internal state of the qubit, $\hat\sigma_z$, to its motion $\hat x$. Thus, the optical Magnus effect allows us to straightforwardly implement a quantum gate.

\emph{Tweezer Hamiltonian --}
In the interaction picture with respect to $\hat H_0=\hbar \omega_m \hat a_m^\dagger \hat a_m$ the tweezer Hamiltonian on ions $i$ and $j$ is:
\begin{equation}\label{eq_Hint}
    \hat H_1=A(t)\left(\frac{1}{2}m\omega_{\rm tw}^2\left(\hat{x}^2_i+\hat{x}_j^2\right)+g\left(\hat{\sigma}_z^{(i)}\hat{x}_i+\hat{\sigma}_z^{(j)}\hat{x}_j\right)\right).
\end{equation}
Here, $\hat{x}_i=\sum_m l_m b_{im}\left(\hat{a}_me^{-i\omega_mt}+\hat{a}_m^{\dag}e^{i\omega_mt}\right)$ is the position operator of ion $i$ in the interaction picture, with $\hat{a}_m^{\dag}$ the creation operator for the mode $m$, and $0\leq A(t)\leq 1$ specifies the time-dependence of the tweezer intensity. The qubit-state independent terms in $\hat H_1$ do not alter the dynamics of the quantum gate. We ignore these terms and arrive at:
\begin{equation}\label{eq_Hint_lin}
\hat H_2=A(t)g\left(\hat{x}_i\hat{\sigma}_z^{(i)}+\hat{x}_j\hat{\sigma}_z^{(j)}\right), 
\end{equation}
which takes the form of a spin-phonon coupling Hamiltonian reminiscent of the M{\o}lmer-S{\o}renson scheme for phonon-mediated quantum gates in trapped ions~\cite{Molmer:1999}. However, at this stage we still have various choices available for $A(t)$, depending on which type of quantum gate we would like to implement. For instance, pulsed $A(t)$ could be used to perform fast gates. Here, we choose $A(t)$ to obtain a geometric phase gate. 
For this, we set $2A(t)=1-\cos (\nu t +\phi)$ where $\phi = 0$ assures a smooth ramp of the tweezer intensity and $\nu=\omega_\text{c}+\delta$ with the subscript $c$ denoting the center-of-mass (c.o.m.) mode for which $\omega_\text{c}=\omega_x$ and $b_{i,\text{c}}=1/\sqrt{N}$. We write the operators $\hat{x}_i$ and $\hat{x}_j$ in terms of $\hat a_c$ and $\hat a_c^\dagger$ and perform the rotating wave approximation to arrive at: 
\begin{equation}\label{eq_Hint_lin_3}
    \hat H_3=\frac{g l_\text{c}}{4\sqrt{N}}\left(\hat{a}_\text{c}e^{i\delta t}+\hat{a}^{\dag}_\text{c}e^{-i\delta t}\right)\left(\hat{\sigma}_z^{(i)}+\hat{\sigma}_z^{(j)}\right).
\end{equation}

To derive the qubit-qubit interactions forming the geometric phase gate, we perform a unitary transformation $\hat U_1=e^{-i\delta \hat{a}^{\dag}_\text{c}\hat{a}_\text{c} t}$ to eliminate the time dependence, followed by a Lang-Firsov~\cite{Lang:1968} transformation, $\hat U_2=\exp \left(\hat{\alpha} \left(\hat{a}^{\dag}_\text{c}-\hat{a}_\text{c}\right)\right)$ with $\hat{\alpha}=-\frac{\tilde{g}}{\delta}\left(\hat{\sigma}_z^{(i)}+\hat{\sigma}_z^{(j)} \right)$. 
Disregarding qubit-independent terms, we obtain
\begin{equation}\label{eq_Heff}
    H_\text{eff}=\frac{2\tilde{g}^2}{\hbar\delta}\hat{\sigma}_z^{(i)}\hat{\sigma}_z^{(j)},
\end{equation}
\noindent with $\tilde{g}=gl_\text{c}/(4\sqrt{N})=\tilde{\eta}\tilde{U}_0$, with the proportionality factor $\tilde{\eta}=\lambdabar l_\text{c}/(\sqrt{N}w_0^2)$. This Hamiltonian generates qubit-qubit interactions that can be used to implement a geometric phase gate by setting the gate time $\tau=2\pi/\delta$ and $\frac{\tilde{g}^2\tau}{\hbar^2\delta}=\pi/4$.

\begin{figure}
    \includegraphics[width=1\columnwidth]{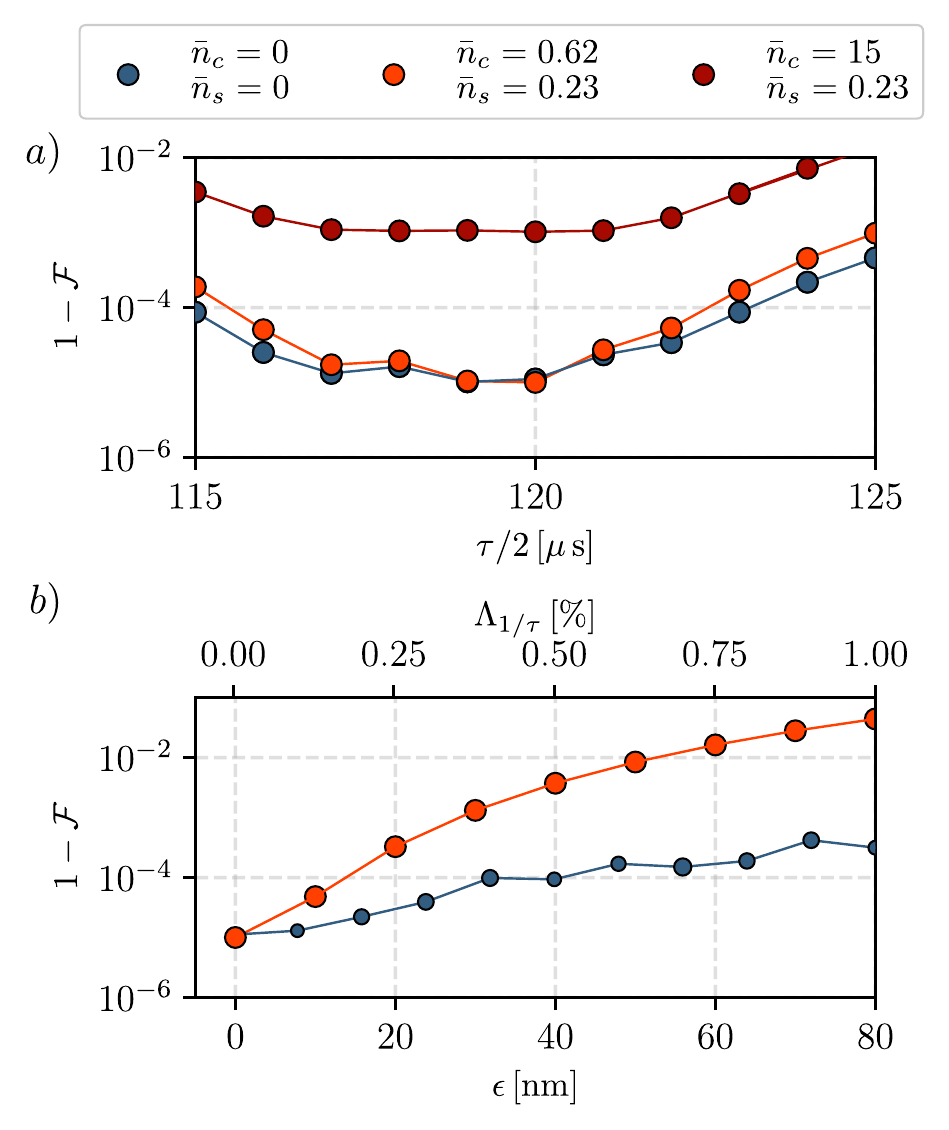}
    \caption{We calculate the gate fidelity for a ground state cooled ion $\bar n_\text{c},\ \bar n_\text{s}=0$ (blue), sub-Doppler cooled thermal state with $\bar n_\text{c}=0.62,\ \bar n_\text{s}=0.23$ (orange) and $\bar n_\text{c}=15,\ \bar n_\text{s}=0.23$ (red, using in this case a Fock cutoff $n_c\leq 120$, $n_s\leq10$). (a) Process fidelity of the two-qubit Magnus gate for different gate times.  (b) Effects of misalignment $\epsilon$ (orange) and intensity noise $\Lambda_{1/\tau}$ (blue) on the gate fidelity. The size of each intensity noise data point represents the standard deviation of 20 simulation where we generated a random Gaussian noise with $\sigma = \Lambda_{1/\tau}$ on each of the two pulses. This implies a noise on the laser intensity at frequency $1/\tau$ that can not be removed by the spin-echo sequence.}
    \label{fig:fidn0}
\end{figure}

\emph{Characterization of the gate -- }
We analyse the gate dynamics by performing numerical simulation of the full dynamics generated by the Hamiltonian  $\hat{H}_{\mathrm{sim}} = \hat{H}_0 + \hat{U}\left(x_i\right)+\hat{U}\left(x_j\right)$ for a two dimensional ion crystal where the tweezers potentials $\hat{U}\left(x_{i;j}\right)$ on ions $i$ and $j$ have been expanded up to fourth-order including spin-independent terms. We use realistic experimental parameters: $\sim$~156~$\mu$W of tweezer laser power focused to a waist of $w_0\sim$~0.5~$\mu$m and tuned 15~THz to the red from the $^2S_{1/2}\rightarrow \, ^2P_{1/2}$ transition in $^{174}$Yb$^+$ ($\lambda=369.5\,$nm). This results in $\tilde{U}_0/h\sim1.6$~MHz, $\tilde{g}/h=2.1$~kHz$/\sqrt{N}$, and setting  $\delta =2\pi\times 12.2$~kHz$/\sqrt{N}$ the gate time for the geometric phase gate is $\tau=170\sqrt{N}\mu$s. With a calculated qubit-state independent tweezer potential of $\omega_{\text{tw}}\sim 2\pi\times 37$~kHz, the center-of-mass mode frequency ($\omega_c/2\pi\sim 1$~MHz) is shifted by $\sim 2\omega_\text{tw}^2/\omega_c N\sim 2\pi\times 710/N$~Hz. This shift can easily be taken into account by correcting $\delta$ accordingly. In these estimates, we neglected the contribution from other dipole allowed transitions, that are detuned by $\sim 66$~THz (the relatively weak $^2S_{1/2}\rightarrow\,^{3}[3/2]_{3/2}$ transition) and 115~THz (the strong D2 line) or more.

We consider the gate unitary with a spin-echo sequence given by
$U(0,\tau)=X^{\otimes 2}U(\tau/2,\tau)X^{\otimes 2}U(0,\tau/2)$,
where $X^{\otimes 2}$ is a qubit flip on both qubits. This spin echo sequence is needed in order to remove local rotations on the qubits states and possible timing errors. We calculate the unitary time evolution operator $U(0,\tau)$ for a system of two ions with their motional c.o.m.\ and \emph{stretch} modes and truncate their respective Hilbert spaces to $n_\text{c}\leq18$ and $n_\text{s}\leq10$. In figure~\ref{fig:fidn0} we show the process fidelity of the gate assuming the ions are in their motional ground state ($\bar n=0$) as a function of gate time.

The gate fidelity of $\mathcal F=$ 0.999988 with $n_\text{c}=n_\text{s}=0$ rivals the current standard approaches. Moreover, the performance of our gate is robust to the thermal occupation of the motional modes. We characterize the gate performance in presence of thermal phonons using the average gate fidelity \cite{Nielsenprocessfidelity, Suppl:Material} and find that it depends weakly on the motional state of the two ions. In fact, using $\bar{n}_\text{c}=0.62$, $\bar{n}_\text{s}=0.23$, the fidelity is almost unaltered at $\mathcal F_{\rm th}=$ 0.999989.

One of main experimental challenges is perfect tweezer alignment. We have studied the resilience of the gate against misalignment of the tweezer in the $x$-direction, which we denote by $\epsilon$. In the presence of misalignment, $\tilde U_0 \to U_0 \exp^{-2 \left(\epsilon + \hat \sigma_z \lambdabar \right)^2/\omega_0^2} $. Thus, the misalignment has two effects: (i) it changes the tweezer potential at the center of the tweezer and therefore the phase accumulation in the phase gate, and (ii) it shifts the potential in a qubit-state-dependent way. The second contribution is corrected to lowest order by a spin-echo sequence. Figure ~\ref{fig:fidn0}(b) shows the infidelity as a function of $\epsilon$. Here we assume that the tweezers are misaligned on both ions in the same way which seems the experimentally most likely case.
The unitary  $U(0,\tau)$ leads to phase space trajectories for $\langle x(t)\rangle$ and $\langle p_x(t)\rangle$ associated with the c.o.m.\ motion\cite{Suppl:Material}. As expected, we find approximately circular phase-space orbits for the even parity states $\ket{00}$, $\ket{11}$, and very little motion for the odd parity ones. We see that every state combination leads to ion motion, but the difference in motion still leads to a high fidelity of $\gtrsim$~0.999 as shown in Figure ~\ref{fig:fidn0}(b).

\begin{figure}
    \includegraphics[width=1\columnwidth]{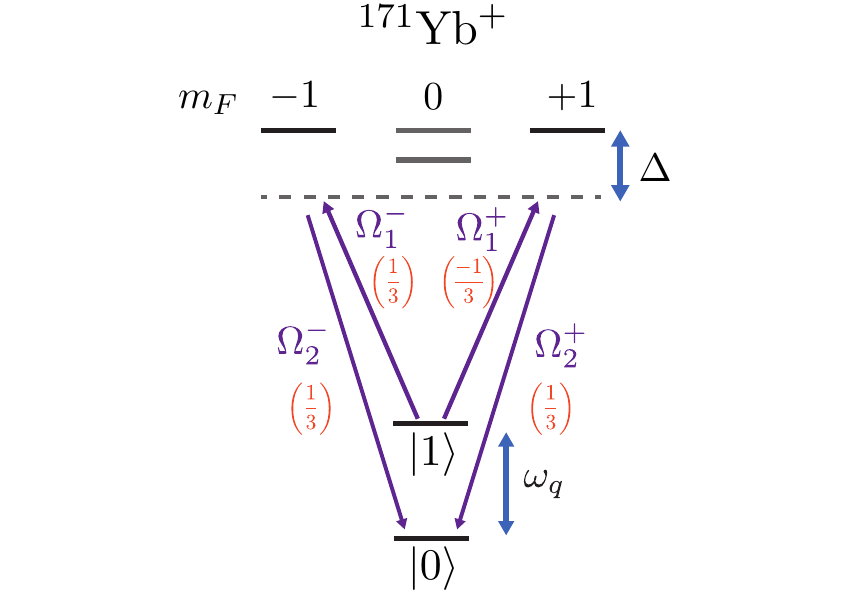}
    \caption{Relevant energy levels of $^{171}\text{Yb}^+$ for implementing the gate on hyperfine qubit splitted by $\omega_q$. The coupling can be achieved using a pair of Raman beams detuned from the upper state $^2P_{1/2}$ by $\Delta$. In the brackets are the angular contributions to the various dipole transition elements.}
    \label{fig:Yb171level}
\end{figure}

\begin{table}
\centering
\renewcommand*{\arraystretch}{1.4}
\caption{Main sources of gate errors. We estimate $\gamma_{\mathrm{ph}}$ as the probability of a off-resonant scattering in for $^{174}$Yb$^+$ during the gate time ($\tau=240\,\mu$s) for a Gaussian and Laguerre-Gaussian beams. 
Other typical sources of errors are misalignment ($\epsilon$), tweezer intensity noise ($\Lambda_{1/\tau}$) and timing ($\Delta\tau$). The values here reported are for laser parameters used in our numerical simulations.}
\resizebox{\columnwidth}{!}{%
\begin{tabular}{|c|c|c|c|c|c|}
\hline
Error source & \begin{tabular}{@{}c@{}}
                   $\gamma_{\mathrm{ph}}$\\
                   Gaussian\\
                 \end{tabular} & \begin{tabular}{@{}c@{}}
                   $\gamma_{\mathrm{ph}}$\\
                   Laguerre-Gaussian\\
                 \end{tabular} & \begin{tabular}{@{}c@{}}
                   $\epsilon$\\
                   $30$\,nm\\
                 \end{tabular} & \begin{tabular}{@{}c@{}}
                   $\Lambda_{1/\tau}$\\
                   0.5\%\\
                 \end{tabular} & \begin{tabular}{@{}c@{}}
                   $\Delta\tau$\\
                   $\pm 5\,\mu\mathrm{s}$\\
                 \end{tabular}\\
\hline
$1-\mathcal{F}$ & $\num{2e-3}$ & $\num{e-6}$ &$\num{1.3e-3}$ &\num{9.3e-5} & \num{2.7e-4}\\
\hline
\end{tabular}
}
\end{table}
\emph{Clock state case -- }
While the calculation was performed for the electron spin qubit states in $^{174}\text{Yb}^+$, it should also be possible to use the hyperfine clock states $\ket{F=m_F=0}$ and $\ket{F=1,m_F=0}$ in $^{171}\text{Yb}^+$. This qubit is insensitive to magnetic field noise and coherence times of up to an hour have been measured~\cite{Wang:2021}. In this case, the tweezers are formed by a bichromatic co-propagating laser field detuned by $\Delta$ from the D1 transition at 369.5~nm with overall detuning $\Delta \ll \omega_\text{FS}$, the fine structure splitting. We set the frequency difference in the bichromatic tweezer to  12.6~GHz, corresponding to the transition between the qubit states~\cite{Olmschenk:2007}. The tweezer laser then causes Raman coupling between the qubit states via two distinct paths. In the first path, the qubits are coupled via the state $|P_{1/2},F=1,m_F=-1\rangle$ due to the $\sigma^-$ polarization component in the tweezer. In the other, the qubits are coupled via the state $|P_{1/2},F=1,m_F=+1\rangle$ due to the $\sigma^+$ component in the tweezer. 
We denote the Rabi frequencies of each path as $\Omega^{\pm}_{1,2}(x)$. The corresponding Raman couplings of each path interfere destructively in the center of the tweezer due to a relative minus sign between $\Omega^+_1(x)$ and $\Omega^+_2(x)$ in their Clebsch-Gordan coefficient,  $\propto(\Omega^{-}_{1}(0)\Omega^{-}_{2}(0)+\Omega^{+}_{1}(0)\Omega^{+}_{2}(0))/\Delta = 0$. 
However, the Magnus effect causes a strong position dependence of the relative strength of both paths of magnitude
\begin{equation}
\Omega_\text{eff}(x)=\frac{\Omega^-_1(x)\Omega^-_2(x)}{\Delta}+\frac{\Omega^+_1(x)\Omega^+_2(x)}{\Delta}\approx \frac{\Omega^2}{\Delta} \frac{4 \lambdabar x}{w_0^2},
\end{equation}
where we assumed $x\ll \lambdabar \ll w_0$ and $|\Omega^\pm_{i}(0)|=\Omega/\sqrt{2}$ with $i=1,2$, such that both laser frequencies have the same power. 
As a result, a qubit state-dependent force appears as in Eq.\ \eqref{eq_Hint_lin}, except that we must now replace $\hat{\sigma}_z^{(i,j)}\rightarrow\hat{\sigma}_x^{(i,j)}$ and the gate takes the form of the usual M{\o}lmer-S{\o}rensen interaction $\propto\hat{\sigma}_x^{(i)}\hat{\sigma}_x^{(j)} $~\cite{Molmer:1999}. Amplitude modulation via $A(t)$ allows again for resonant enhancement of the gate.

In addition to the Raman coupling, we obtain a tweezer potential (AC Stark shift) for each qubit state of magnitude
\begin{equation}
\delta^{|k\rangle}_\text{AC}(x)=\sum_{i=1,2}\sum_{j=+,-}\frac{|\Omega^j_i(x)|^2}{\Delta_{i,|k\rangle}}
\end{equation}
with $\Delta_{1,|0\rangle}=\Delta-\omega_q$, $\Delta_{2,|0\rangle}=\Delta$, $\Delta_{1,|1\rangle}=\Delta$ and $\Delta_{2,|1\rangle}=\Delta+\omega_q$.
This causes an additional trapping potential $\Phi(x)\approx\frac{1}{2}m\omega_\text{tw}^2x^2$ that is independent of the qubit state as before, as well as a position-dependent differential Stark shift $\delta_\text{AC}(x)=\delta^{|1\rangle}_\text{AC}(x)-\delta^{|0\rangle}_\text{AC}(x)$. 
In the limit $\omega_q\ll|\Delta|$, 
\begin{align}
        \delta_\text{AC}(x) &\approx -\frac{\omega_q}{\Delta^2}\sum_{i=1,2}\sum_{j=+,-}|\Omega^j_i(x)|^2\\
        & = -\frac{\omega_q}{\Delta}\,\tilde{U}_0(x)
\end{align}
This differential Stark shift is estimated to be small, $\delta_\text{AC}/2\pi\approx 2.7\,$kHz for the numbers used in the simulations, and can be compensated by a corresponding Raman detuning.

Photon scattering on the D1 transition can be estimated as $\gamma_\text{ph}\sim \tilde{U}_0\Gamma/(\hbar\Delta)\sim 13$~s$^{-1}$ with $\Gamma=1.23 \times 10^8$~s$^{-1}$ in Yb$^+$. This adverse effect may be reduced significantly by employing hollow tweezers~\cite{Schmiegelow:2016,Drechsler:2021,Mazzanti:2021} at the expense of added complexity. For a hollow beam with a waist $w_0 = 0.5\,\mu$m and $\sim 160\,\mu$W we obtain a reduction in scattering rate of $\sim 10^{-6}$~s$^{-1}$. As long as $\omega_\text{tw}\ll \Omega_\text{rf}$, the drive frequency of the Paul trap, no parametric excitations can occur and micromotion of the ions is not a problem. Other errors, such as due to intensity noise of the laser, heating of the ions due to electric field noise and decoherence due to magnetic field noise have the same effect as in other gate implementations. Finally, we note that because the tweezers are far detuned from the closest transitions, the exact overall frequency of the tweezer laser is irrelevant.

\emph{Conclusions} 
In conclusion, we have described a novel type of quantum phase gate based on the optical Magnus effect using optical tweezers in a linear chain of trapped ions. The main benefit is that the gate does not require counter-propagating laser fields, greatly simplifying the setup and eliminating errors due to phase instabilities between the gate laser beams. Furthermore, the state-dependent force generated by the Magnus effect allows to perform the gate by coupling to motional modes on the plane perpendicular to the direction of propagation of the tweezers allowing novel experimental implementations. The proposed gate does not require ground state cooling and can perform a quantum logic gate on any pair of ion qubits by spatial addressing. The expected gate fidelity rivals the state of the art also for ions that are not cooled to the ground-state of motion.

\section*{Acknowledgements}
This work was supported by the Netherlands Organization for Scientific Research (Grant Nos. 680.91.120 and 680.92.18.05, (R.G.). A.S.N is supported by the Dutch Research Council (NWO/OCW), as part of the Quantum Software Consortium programme (project number 024.003.037).

\bibliographystyle{apsrev4-2}

\bibliography{biblio-RG,magnusgate}

\thispagestyle{empty}
\addtocounter{page}{-1}

\clearpage

\setcounter{equation}{0}
\setcounter{figure}{0}
\setcounter{table}{0}
\setcounter{page}{1}

\counterwithin{figure}{section}
\renewcommand\thefigure{\arabic{figure}}

\let\theequationWithoutS\theequation 
\renewcommand\theequation{S-\theequationWithoutS}
\let\thefigureWithoutS\thefigure 
\renewcommand\thefigure{S-\thefigureWithoutS}

\renewcommand*{\citenumfont}[1]{S#1}
\renewcommand*{\bibnumfmt}[1]{[S#1]}

\title{Supplementary material for : \\Trapped Ion Quantum Computing using Optical Tweezers and the Magnus Effect}

\maketitle

\section*{Appendix I : Optical Magnus effect}
\begin{figure}[h!]
    \includegraphics[width=\columnwidth]{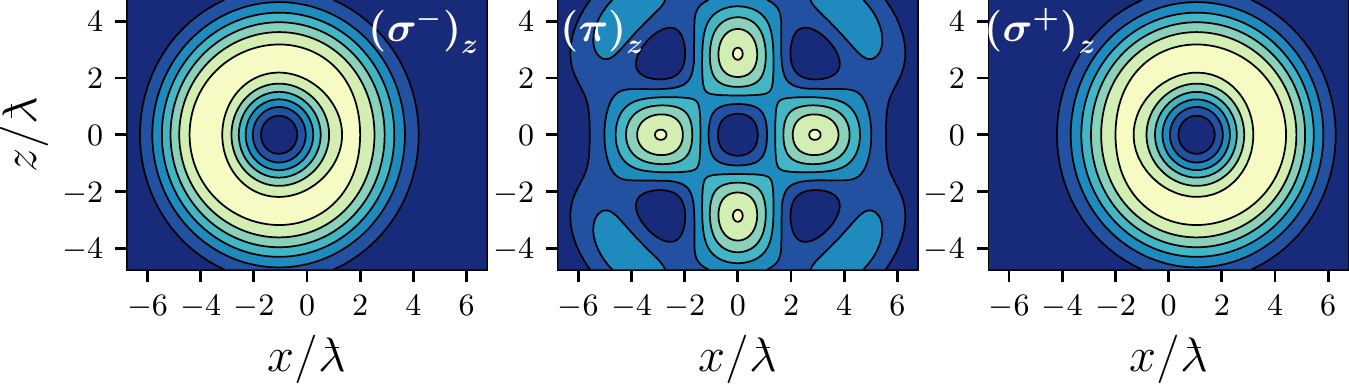}
    \caption{Intensity of the polarization components for a LG beam calculated at the focus.  The $\pi$-polarization component has been enhanced by a factor $100$ to make it visible. Here we set $w_\theta = 0.6$}
    \label{fig:LGcomponents}
\end{figure}

A key characteristic of a tightly focused beam is the strong field curvature near the focus. This not only affects the local intensity but also its polarization structure. To calculate this, we take a superposition of plane waves labeled by their wave vector in spherical coordinates,  $\mathbf{k}=(k,\theta,\phi)$. Taking $k=\omega/c$ as fixed we write
\[\mathbf{E}(\mathbf{r})\propto\int_0^{2\pi} d\phi\int_0^\pi d\theta\,\sin\theta\, \mathbf{u}_x(\theta,\phi)\,a(\theta,\phi)\, e^{i\mathbf{k}\cdot\mathbf{r}}\]
with $\mathbf{u}_x(\theta,\phi)$ a polarization vector obtained by co-rotating the $\mathbf{x}$ unit vector when $\mathbf{k}$ is rotated from $\mathbf{z}$ to $(\theta,\phi)$, such that $\mathbf{u}_x(\theta,\phi)\cdot\mathbf{k}=0$, see also Ref.\ \cite{spreeuw_off-axis_20202}.
In the calculation we center the beam around $\theta=0$, and the focal plane is given by $\mathbf{r}=(x,y,0)$. The shape of the beam is determined by the amplitude function $a(\theta,\phi)$. For a Gaussian beam we set $a(\theta,\phi)=\exp(-\theta^2/w_\theta^2)$; for the lowest order ($l=1$) Laguerre-Gaussian (LG) beam we set
$a(\theta,\phi)=\theta\exp(i\phi-\theta^2/w_\theta^2)$.
After performing the above integral we rotate the results for tweezers propagating along the $-y$ direction. Finally, the circular field components $\sigma^\pm$ shown in Fig.\ 1 of the main text are obtained as the projection onto unit vectors $(\mathbf{x}\pm i\mathbf{y})/\sqrt{2}$. In Figure \ref{fig:LGcomponents}, all three polarization components for a Laguerre-Gaussian beam are shown. Note that the $\sigma^-$ and $\sigma^+$ components have similar intensity while the $\pi$-polarization is suppressed by a factor $\sim 100$.
\section*{Appendix II : Phase-space dynamics}
\begin{figure}[h!]
    \includegraphics[width=\columnwidth]{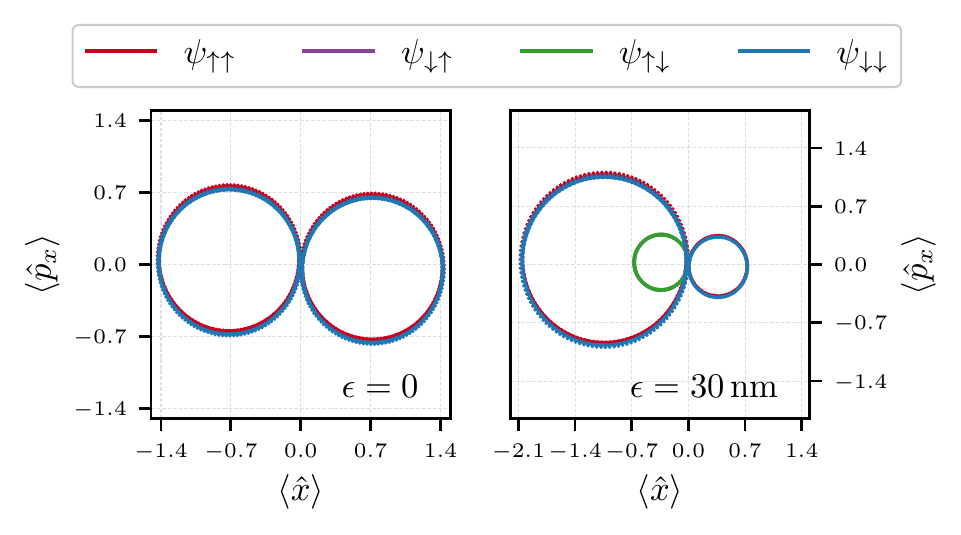}
    \caption{Center of mass mode phase-space dynamics for perfectly aligned tweezer (left) and for $30$\,nm misaligned ones (right). For the simulation we used the same parameters as for $\tau/2 = 120 \,\mu$s point in Figure 1(a) of the main text.\\}
    \label{fig:phsp}
\end{figure}

We study the phase-space dynamics of the ions by simulating the time dependent Hamiltonian using trotterization with time-steps of $10^{-4}\,\tau$. At each time-step we evaluate the expectation value of the $\langle\hat{x}\rangle$ and $\langle\hat{p}\rangle$ for the center of mass mode. As expected, we find approximately circular phase-space orbits for the even parity states $\ket{00}$, $\ket{11}$, and very little motion for the odd parity ones. In Fig. \ref{fig:phsp} it is possible to see the evolution in phase-space for all the four spin states in case of perfectly aligned and slightly misaligned tweezers. As described in the main text we simulate numerically the full Hamiltonian defined as $\hat{H}_{\mathrm{sim}} = \hat{H}_0 + \hat{U}\left(x_i\right)+\hat{U}\left(x_j\right)$ where in case of misalignment $\epsilon$, $ \hat{U}\left(x\right)$ reads as :

\begin{align}
&U(x)\approx-U_0\, e^{-2(\left(\hat{x}-\hat{\epsilon} \right)+\hat{\sigma}_z\lambdabar)^2/w_0^2}\nonumber\\
&\approx -\tilde{U}_0 + 4\tilde{U}_0 \frac{ \hat{\sigma}_z\lambdabar - \hat{\epsilon} }{w_0^2}\hat{x} \nonumber\\&
+ \frac{1}{2}\tilde{U}_0\left(\frac{4\left(w_0^2 - 4\lambdabar^2\right)}{w_0^4}\right)\hat{x}^2 -\frac{1}{2}\tilde{U}_0\left( \frac{16\left(\hat{\epsilon}^2-2\hat{\sigma}_z\hat{\epsilon}\lambdabar\right)}{w_0^4}\right)\hat{x}^2 \nonumber\\&
-\left(8\tilde{U}_0\hat{\sigma}_z\lambdabar \frac{3w_0^2-4\left(3\hat{\epsilon}^2+\lambdabar^2\right)}{3w_0^6} \right)\hat{x}^3\nonumber\\&
+\left(8\tilde{U}_0\hat{\epsilon} \frac{3w_0^2-4\left(\hat{\epsilon}^2+3\lambdabar^2\right)}{3w_0^6} \right)\hat{x}^3 \nonumber\\&
-\left(
2\tilde{U}_0\hat{\sigma}_z\lambdabar\hat{\epsilon} \frac{-48w_0^2+64\left(\hat{\epsilon}^2+\lambdabar^2\right)}{3w_0^8}\right)\hat{x}^4 \nonumber\\&
+\left(2\tilde{U}_0 \frac{3w_0^4-24w_0^2\left(\hat{\epsilon}^2+\lambdabar^2\right)+16\left(\hat{\epsilon}^4 + 6\hat{\epsilon}^2\lambdabar^2+\lambdabar^4\right)}{3w_0^8}\right)\hat{x}^4.\nonumber
	\end{align}
with 
\begin{align}
	\tilde{U}_0 = U_0 e^{-2\left(\hat{\epsilon}+\hat{\sigma}_z \lambdabar\right)^2/w_0^2}\nonumber
\end{align}
A small tweezer misalignment $\epsilon$ gives rise to new spin-dependent terms in the Hamiltonian that shift the trapping potential in a state dependent way. In Fig.\ref{fig:phsp} is shown how the dynamics is affected in the case where the tweezers are misaligned by 30~nm.
\section*{Appendix III : Gate fidelity}
We characterize the gate by calculating the average process fidelity as follows : \cite{Nielsenprocessfidelity2}:
\begin{equation}
\bar{F}(\hat{U}_{\text{id}},\hat{U}_{\hat{H}_{\mathrm{sim}}})=\frac{\sum_j \text{tr}\left[\hat{U}_{\text{id}}\hat{\sigma}_j^\dagger\hat{U}_{\text{id}}^\dagger\boldsymbol{\hat{\sigma}}_j(\hat{U}_{\hat{H}_{\mathrm{sim}}})\right]+d^2}{d^2\left(d+1\right)},\nonumber
\end{equation}
where $\hat{U}_{\text{id}}$ is the unitary of an ideal geometric phase gate and $\boldsymbol{\hat{\sigma}}_j(\hat{U}_{\hat{H}_{\mathrm{sim}}}) \equiv \text{tr}_{\text{FS}}(\hat{U}_{\hat{H}_{\mathrm{sim}}}\left[|n\rangle\langle n|\bigotimes\hat{\sigma}_j\right]\hat{U}^\dagger_{\hat{H}_{\mathrm{sim}}})$ projects the unitary matrix generated by the time evolution of the Hamiltonian used for the simulations $\hat{U}_{\hat{H}_{\mathrm{sim}}}$ on the Fock state $|n\rangle$ and on a $d$-dimensional representation Pauli matrices.

\end{document}